\documentstyle[aps, multicol]{revtex}
\begin{document}
\newcommand{\be}{\begin{equation}}
\newcommand{\ee}{\end{equation}}
\newcommand{\bq}{\begin{eqnarray}}
\newcommand{\eq}{\end{eqnarray}}
%
\title{String Limit of Vortex Current Algebra} 
\vspace{1. truecm} 
\author{Vittorio Penna$^{1}$, and Mauro Spera $^{2}$}
\vspace{.25 truecm}
\address{$^{1}$ Dipartimento di Fisica and Unit\`a INFM,
Politecnico di Torino, C.so Duca degli Abruzzi 24, I-10129 Torino,
Italy \\
and ESI, Boltzmanngasse 9, A-1090, Wien, Austria}
\address{ $^{2}$ Dipartimento di Metodi e Modelli Matematici, 
Universit\`a di Padova, I-35131 Padova, Italy}
\date{\today}
\maketitle
\begin{abstract}
The Poisson structure generating the hamiltonian dynamics of string
vortices is reconstructed within the current algebra picture as a
limiting case of the standard brackets associated to fluids with a
smooth vorticity field. The approach implemented bypasses the use
of Dirac's procedure. The fine structure of the dynamical algebra
is derived for planar fluids by implementing an appropriate spatial
fragmentation of the vorticity field, and the limit to the point
vortex gas is effected. The physical interpretation of the resulting
local currents is provided. Nontrivial differences characterizing
the canonical quantization of point vortices and the current algebra
quantization are also illustrated.
\end{abstract}
\begin{multicols}{2}
\section{Introduction}
The formation of vortices and their interactions in superfluid
media~\cite{DON} such as $^{4}$He (and closely related systems
such as type-II superconductors~\cite{BG2})
have been detected and thoroughly studied at
the classical level since forty years ago.
Recent experimental results concerning the Bose-Einstein condensates
show the emergence of vortices in the condensates~\cite{BEC} thus
providing a further scenario in which vortices can be investigated.
Despite the large number of physical systems exhibiting excited states
with vortices, a quite mild interest has been raised by
the study of the quantum aspects inherent in their dynamics that,
on the contrary, should be relevant both because vortex formation
occurs at very low temperatures, and because vortex interactions
take place on microscopic/mesoscopic spatial scales, where quantum
effects are important~\cite{NEW}.

Such a situation is probably originated by the great difficulties
in formulating, within the quantum field theory of superfluid media
(and closely related systems), a quantization scheme which supplies
both an effective representation of the fluid topological excitations
(vortex states), and a viable approach to investigate the formal
aspects of the theory.
In particular, the dynamical degrees of freedom activated by the
vortex emergence exhibit a structural complexity which renders
dramatically difficult any attempt to construct explicitly
the Hilbert space for the fluid quantum states
\cite{RR}, \cite{GMS1}.
Such a program is further complicated by the fact that, since vortices
are extended object equipped with a possibly nontrivial topological
structure, a consistent quantum description of the fluid should
incorporate as well the vortex topology in terms of constant of
motions representing generalized circulations.

Such aspects have been thoroughly studied in a series
of paper within the geometric quantization scheme both for 
fluids with a vorticity field confined on filaments \cite{GMS1},
\cite{GMS2}, \cite{PS1} (gas of line vortices), and for
fluids whose state is described by a smooth (extended) vorticity
field \cite{PRS}, \cite{WU}. A large amount of work has been
devoted to such two models of fluids in order to realize the
unitary irreducible representations of the field operators and
the ensuing construction of the Hilbert space.
Despite the recognition within the geometric scheme
of several basic, both group-theoretic and algebraic,
features that characterize the fluid structure and its
description, almost no attention has been directed to
considering how the vortex quantization is influenced by
the limiting process which relates the previous models
through the squeezing of the extended vorticity field
to a set of disjoint lines.

In this paper we start to investigate this limit and
the advantages it entails as to the stringlike-vortex
description, and try to emphasize some nontrivial aspects
concerning the quantization of vorticity fields in the
planar case.
 
One of the first attempts to quantize the vortex dynamics
was developed in Ref. \cite{FET} for a model of almost
parallel line vortices within the canonical approach based
on coordinates and momenta. 
Its natural extention to planar systems of superfluids with point 
vortices raised a certain interest several years later mainly in
relation to the possibility of observing fractional statistics. 
The difficulties inherent in the
quantization process were completely recognized in Ref. \cite{RR}
where the canonical scheme was employed to construct the quantum field 
theory of three-dimensional (3D) vortices characterized
by a singular vorticity field
\be
{\bf w}({\bf x}) =
\, k \,\oint_{\Gamma} d{\bf x}(s) \,\, \delta^3 ({\bf x}-{\bf x}(s))
\label{VOR}
\ee
with vortex strength $k$, where $\Gamma$ is a possibly self-knotted
string with any number of components. Such an arbitrarily complex
object provided a realistic generalization of the model of parallel
vortex lines by introducing explicitly the topological strucure of
line vortices. 
The components $x_j(s,t)$ of the 3D vector ${\bf x}(s,t)$ representing
the smooth curve $\Gamma \in {\bf R}^3$ ($s$ is the string parameter
on $\Gamma$, and $j=1,2,3$) supplied the coordinates at each time $t$,
whereas the canonically conjugate momenta
$$
P_i(s,t) := \quad
\frac{\delta {\cal L}}{\delta {(\partial_t  x_i ) } } 
$$
were obtained from the Lagrangian functional
$$
{\cal L} := \; - H \, + 
\, \frac{k \rho}{3} \, \oint_{\Gamma} d{\bf x} \,\cdot 
\Bigl ( \, \frac{ \partial {\bf x} }{\partial t}  \wedge {\bf x} \, 
\Bigr ) \; ,
$$
(the fluid density $\rho$ is assumed to be constant)
containing the ideal fluid energy $H$ (see Eq. (\ref{IFH})
below). Momenta $P_j$ entail the dynamical constraints
$ P_i \! -\! (\rho k/3) \, \epsilon_{ijk } 
\, x_j \, \partial_s x_k = 0 $
that revealed the singular character of $\cal L$, and showed
how the stringlike vortex dynamics actually takes place on a
submanifold of the standard phase space
${\cal P} := \{ (P_j(s,t), x_j(s,t) ) \}$.
The price of implementing the canonical picture based on the brackets
\be
\{ x_i(s ,t) , P_j(s' ,t) \} = \, \delta (s-s') \, \delta_{i,j}
\label{CBR}
\ee
was to reconstruct Eq. (\ref{CBR}) within the Dirac procedure so as to
incorporate the dynamical constraints. The dynamics was thus formulated
through the Dirac brackets
$\{ A,B \}_* := $$ \{ A,B \} + \langle A|C| B \rangle$, where
$$
\langle A|C| B \rangle 
\equiv \int \! ds  \! \int \! \! ds' \! 
\{ A ,\Phi_\alpha (s) \} \, C_{\alpha, \beta}(s,s')
\,  \{ \Phi_{\beta} (s'), B \} \, ,
$$
$\{ A , B \}$ represents the standard Poisson brackets,
$C_{\alpha,\beta}$ are elements of the matrix 
$C := ||\{ \Phi_{\alpha},\Phi_{\beta} \}||^{-1}$ and
the functionals $\Phi_{\alpha}(s)$, $\alpha =1,2$, of the canonical
variables essentially identify with the components of the part of 
${\bf P} -(\rho k/3) {\bf x} \wedge \partial_s {\bf x}$ orthogonal to
the vector field ${\bf x}(s,t)$. The main issue of Dirac's formalism 
was the unexpected coordinate brackets 
\be
\{ x_i(s) , x_j(s') \}_* \, = \, \frac{1}{k \rho}
\epsilon_{ijk} \, \delta(s-s') \, \partial_s x_k(s)
\label{SBR}
\ee 
showing how coordinates $x_j$ cannot be regarded any longer as
independent variables. At the quantum level, Eqs. (\ref{SBR})
entailed the remarkable effects that the projections of
$\Gamma$ on the planes $x_i-x_j$, $i, j= 1,2,3$, are affected
manifestly by the quantum uncertainty, and led to imagine
$\Gamma$ as a tubular domain representing the intrinsically
approximate position of the vortex core (see Ref. \cite{BIGAS}).

A further observation is suggested by Eq. (\ref{SBR}): in spite
of the local canonical character of coordinates, the algebraic
structure exhibited by Eq. (\ref{SBR}) is actually nonlocal
consistent with the fact that $\Gamma$ is a true
three-dimensional object.
In Ref. \cite{RR}, this led the authors to construct the algebra of
currents, which we review in the sequel, so as to avoid the
dependence on local parametrizations as well as on Dirac's formalism.

The first goal of this paper is to show how the algebraic structure
involved by the functional picture based on 
$\Gamma$-dependent currents
can be derived in a direct way from the standard Lie-Poisson structure
\be
\{ F, G \}({\bf w}) = {1 \over \rho} \int d^3x \,\,
{\bf w} \cdot \Bigl( {\rm curl}\, {{\delta F} \over {\delta {\bf w}}}
\wedge {\rm curl}\, {{\delta G} \over {\delta {\bf w}}} \Bigr)\, ,
\label{PBW}
\ee
without implementing Dirac's procedure. The structure
(\ref{PBW}), that generates the vortex dynamics~\cite{KUMI}
when the vorticity field ${\bf w}({\bf x})$ is smooth (namely its
components ${\rm w}_j({\bf x}) \in C^{\infty}({\bf R}^3)$),
contains as a limiting case the Poisson structure for vorticity
fields $\bf w$ collapsed on an array of strings (singular limit),
no matter how complex the underlying topological structure is.

The second purpose of the present paper is to consider the effect
of the singular limit on the vortex Lie-Poisson (LP) structure in
the two-dimensional case, where the limit consists in squeezing
the vorticity fields on (a set of) isolated points of the
ambient plane.
The resulting point vortex gas is well known to represent the 
reference model for a number of systems with vortex excitations
such as superfluid films \cite{BG1} of $^4$He (adsorbed both on
planar substrates and on porous materials), planar superconductors
\cite{BG2}, and Josephson junctions' arrays \cite{BG3} (see also
Ref. \cite{PV} and references therein).
Our interest in analyzing the 2D singular limit comes from the
wish of establishing a clear link between the smooth case and
the singular case. More specifically, we aim at unveiling the
algebraic structure of point vortices within the ampler
framework of the current algebra (CA) of the smooth case.

The point of view adopted here is that in the 2D case the CA
contains an explicit many-body structure related to the
spatial distribution of positive/negative vorticity which
deserves to be investigated. Such a fine structure of
$\bf {\cal A}$ (storing information on the spatial distribution
of $\bf w$) paves the way to the emergence of the canonical
Poisson structure that customarily characterizes the point
vortex dynamics.
The 3D case can be also studied from this viewpoint even if the
fragmentation must be developed based on the complex topological
structure of $\bf w$. In this respect, however, the Arnold cells
\cite{ARN} should represent the 3D counterpart of 2D fragmentation.
The latter enables us to shed light on certain features that
characterize, at the quantum level, the construction of the CA
of the point vortex model, and unexpectedly prevents it from
matching the version of the algebra obtained within the field
theory formalism.

The paper is organized as follows: In Sec. II, after
introducing the standard derivation of Euler's equations
by means of appropriate Lie-Poisson brackets, the CA
picture is rewieved for the 3D case and its relevance
for the quantization of the vortex dynamics is showed.
In Sec. III the CA picture is used to perform in a
consistent way the string limit of formula (\ref{PBW})
in order to construct the Poisson structure for a gas of
stringlike objects (a more formal derivation is furnished
in appendix A based on the Clebsch potential picture).
Some applications of the brackets thus obtained are illustrated
as well. In Sec. IV the many-body structure of vortex dynamics
is investigated in the 2D case and the fine structure of the CA
is evidenced via an appropriate fragmentation of the vorticity
field. The latter is related to the CA reconstructed for 2D point
vortices within the canonical quantization. The CA of the point
vortices is compared with the CA of the smooth case and their
inequivalence at the quantum level is proven: a suitable
semiclassical limit is shown to reconcile such situations.

\section{Current Algebra approach to Vortex dynamics}
The classical motion of a perfect fluid with velocity field $\bf v$,
vorticity field ${\bf w} = {\rm curl}\, {\bf v}$, and Hamiltonian
\be
H[{\bf v}] = {\rho \over 2}
\int_{\bf R}^3 \! d^3x \,\,
{\bf v}^2(\bf x) \; ,
\label{IFH}
\ee
is governed by the Euler equation
$
\dot {\bf v} = - {\bf v} \cdot \nabla \,\, {\bf v}.
$
Observing that ${\bf v} \cdot \nabla \, {\bf v} = 
{\bf v} \wedge {\bf w} -\! \nabla {\bf v}^2/2$,
the vorticity equation
\be
\dot {\bf w} = - {\rm curl}\,( {\bf v} \wedge {\bf w} ) \,\, ,
\label{WW}
\ee
based on representing the fluid state through $\bf w$, is easily
obtained from the Euler equation \cite{ARK}.
The derivation of both equations is easily
performed by means of the usual Lie-Poisson (LP) brackets \cite{MWA}
\be
\{ F ,G \}[{\bf v}] =
{1 \over \rho} \int d^3x \,\, {\rm curl}\, {\bf v} \cdot 
\Bigl( {{\delta F} \over {\delta {\bf v}}} \wedge 
{{\delta G} \over {\delta {\bf v}}} \Bigr)
\label{PBV}
\ee
where $F$ and $G$ are functions that depend on $\bf v$, and
the notation $\int d^nx $ denotes, from now on, the integration
on the whole space ${\bf R}^n$. It is
important to observe how a consistent use of Eq. (\ref{WW})
requires that $\bf v$ is expressed as a functionals of $\bf w$.
This is achieved by imposing the divergenceless condition
${\rm div }{\bf v} =0$ on $\bf v$, namely by identifying $\bf v$ with
$ {\bf V(x) } = {\rm curl} \, {\bf U(x)}$,
where the vector potential {\bf U(x)} is defined as
$$
{\bf U(x)} 
= \int d^3y \,\, G({\bf {x - y}})\, {\bf w(y)} \,\, ,
$$
and the Green function $G$ in 3 and 2 dimensions reads 
$G({\bf x}\! - \!{\bf y}) \! = \! 1/(4\pi \vert {\bf x}\! - \!{\bf y}
\vert)$ and
$G({\bf x}\! - \!{\bf y} ) = \!(1/2 \pi ){\sl ln}
\vert {\bf x}\! -\! {\bf y} \vert$, respectively.

In the case when $F$ and $G$ are assumed to depend on the
equivalence class $[{\bf v}] = \{ {\bf v'}: {\bf v'} =
{\bf v } + \nabla f \}$,
the LP brackets take the form (\ref{PBW}) in that
${\delta F}/\delta {\bf v}={\rm curl}\,({\delta F}/\delta{\bf w})$.
Explicitly, this amounts to assuming that $F$ and $G$ depend on
$\bf w$ namely on the divergence-free field $\bf V$.

Replacing $\bf v$ with $\bf V$ in $H[{\bf v}]$ incorporates
explicitly the divergenceless feature in the theory. This leads
to the quadratic form 
$$
H[{\bf w}] \equiv 
{\rho \over 2} \int \! d^3x \int \! d^3y \,\, 
G({\bf x - y}) \, {\bf w}({\bf x}) \cdot {\bf w}({\bf y}) \,\, ,
$$
which generates Eq. (\ref{WW}) via brackets (\ref{PBW}).

An alternative formulation~\cite{RR} of vortex dynamics
can be given in terms of current algebra $\cal A$. The latter
consists of functionals of $[{\bf v}]$ (the currents) defined as  
\be
J_{\bf a} [{\bf v}]= \rho \int \! d^3x \,{\bf a} \! \cdot \!{\bf v} 
= \rho \int \! d^3x \,{\bf A} \cdot {\bf w} = J_{\bf A}[{\bf w}] \, ,
\label{CURR}
\ee
where $\bf a$ belongs to the algebra
${\cal G} = \{ {\bf a}: {\bf a} = {\rm curl} \,{\bf A} \}$
of divergence-free vector fields. One can easily check that
$J_{\bf a}[{\bf v}] =J_{\bf a}[{\bf v} + \nabla f]$.
The algebraic structure of $\cal A$
shows up via the equation \cite{RR2}
\be
\{ J_{\bf a} ,J_{\bf b} \} [{\bf v}] =
J_{\bf {[a,b]}} [{\bf v}] \,\,\, ,
\label{CALG}
\ee
where ${\bf [a,b]} ={\rm curl}({\bf a \wedge b })$, that is fulfilled
by two any currents of $\cal A$. The structure constants of $\cal A$
are readily worked out by introducing the subalgebra
of the mode currents ${\cal A}_F$. The latter is defined by
noting that any current $J_{\bf A}[{\bf w}]$ can be expressed
in terms of the Fourier transform ${\bf A }({\bf q})$ relative
to $\bf A$ as
$$
J_{\bf A}[{\bf w}] = \rho \int \! d^3x \,\,{\bf A} \cdot {\bf w}
= \int \! d^3q \,\,{\bf A}({\bf q}) \cdot {\bf J}_{\bf q}[{\bf w}]\, ,
$$
where the mode current
${\bf J}_{\bf q} [{\bf w}] := {\bf e}_m J_{\bf q}^m [{\bf w}]$
has vector components
\be
J_{\bf q}^m [{\bf w}] =
\rho \int\! d^3x \,\,e^{i{\bf x}{\bf q}} {\sl w}_m({\bf x}) \; .
\label{TRA}
\ee
Then one can easily show that the basic brackets of $\cal A$
are given by
\be
\{ J_{\bf q}^m ,J_{\bf p}^n \} [{\bf w}] =
- \sum_{k=1}^3 \,
C_{m,n}({\bf q}, {\bf p})
\,J_{\bf q + p}^k [{\bf w}] \, ,
\label{MALG}
\ee
where
$C_{m,n}({\bf q}, {\bf p}) := {\bf e}_k \cdot [({\bf q}\wedge 
{\bf e}_m )\wedge ({\bf p}\wedge {\bf e}_n)]$ are the structure
constants, and ${\bf e}_k$, $k=1,2,3$ are the unit vectors of the
3D euclidean basis. The equations of motion of any current
$J_{\bf q}^m [{\bf w}]$ can be easily derived via the brackets
(\ref{MALG}) once $H[{\bf w}]$ itself has been rewritten as a
functional of currents $J_{\bf q}^m [{\bf w}]$ by inverting
formula (\ref{TRA}) (see Ref. \cite{RR2}).
This proves that ${\cal A}_F$
indeed furnishes a set of observables which is complete, namely
${\cal A}_F$ represents an alternative scheme in which representing
the fluid dynamics. Brackets (\ref{MALG}) also shows how the
sub-algebra of mode currents is the most advantageous set of
commutators in which implementing the quantization process.
Nevertheless, it important to recall that the parent set of
commutators (\ref{CALG}) indicate that the quantization
process is equivalent to construct the unitary irreducible
representations of the group of diffeomorphism which is
one of the hardest, unsolved problem of the theory of
group representations.
\section{String limit of 3d Current Algebra}

Upon performing the limit which confines the vorticity field on a
stringlike domain (that is, on a vortex filament), namely
considering
\be
{\bf w}({\bf x}) \to {\bf w}_*({\bf x}) \equiv
\sum_a k_a \, \oint_{\Gamma_a} \! d{\bf y}_a(s) \,\,
\delta^3 ({\bf x} - \! {\bf y}_a(s))
\label{LIM}
\ee
the current (\ref{CURR}) becomes
$$
J_{\bf A} [{\bf w}]
= \rho \sum_a k_a \,\oint_{\Gamma_a}\! d{\bf x}_a \cdot
{\bf A}({\bf x}_a) \, ,
$$
whereas the current $\{ J_A ,J_B \}[{\bf w}]$ reduces to
$$
\{ J_A ,J_B \}[{\bf w}] =
\rho \sum_a k_a \,\oint_{\Gamma_a} \!\! d{\bf x}_a \cdot
({\bf a}({\bf x}_a) \wedge {\bf b}({\bf x}_a)) \quad.
$$
Therefore the form assumed by Poisson brackets (\ref{PBW})
when limit (\ref{LIM}) is enacted must be consistent with this
result. Observing that
\be
{{\delta {\cal F} } \over {\delta {\bf x}(s)}} =
{{\partial f } \over {\partial {\bf x}(s)}} -
{d \over {ds}} {{\partial f } \over {\partial {\bf \dot x}(s)}} 
\label{DER}
\ee
for any function 
${\cal F} = \int dq \,\,f[ {\bf x}(q),{\bf \dot x}(q) ]$, where
${\bf \dot x}(s) := {{d{\bf x}(s) }/{ds}}$, one easily obtains
$$
{{\delta J_{\bf A}} \over {\delta {\bf x}(s)}} = \, \rho k
\,\,\,{\bf \dot x}(s) \wedge {\rm curl}\,{\bf A} \, \, . 
$$
This entails, in turn
$$
{\bf \tau}(s)\cdot \left [
{{\delta J_{\bf A} } \over {\delta {\bf x}(s)}}
\wedge
{{\delta J_{\bf B} } \over {\delta {\bf x}(s)}} \right ]
=\, k^2 \rho^2 \,
{\bf \tau}(s) \cdot ( {\bf a} \wedge {\bf b})_{\bf x(s)} 
$$
provided ${\bf \tau}(s) ={\bf \dot x}(s)$ is a unit vector $\sl i.e.$
the parameter $s$ is identified with the arc-lenght of $\Gamma$.
This result suggests the substitution
\be
{\rm curl}\, {{\delta F} \over {\delta {\bf w}}} \rightarrow
{\bf \dot x}_a \wedge
{{\delta } \over {\delta {\bf x}_a}} \frac{\partial F}{\partial k_a}
\label{DWLIM}
\ee
as a consequence of limit (\ref{LIM}), where the index $a$ takes into
account the possible many-component structure $\Gamma =\{ \Gamma_a \}$
of the string model. Hence the string LP brackets for a many-component
line vortex turn out to be
$$ 
\{ F ,G \}[{\bf w}] =  \sum_a  
\oint_{\Gamma_a} \! \frac{k_a d{\bf x}_a}{\rho} \cdot 
\left ( 
{{\delta } \over {\delta {\bf x}_a}} \frac{\partial F}{\partial k_a} \wedge 
{{\delta} \over {\delta {\bf x}_a}} \frac{\partial G}{\partial k_a}
\right ) \, .
\label{SLPBG}
$$
A simplified version is also available in the form 
\be
\{ F ,G \}[{\bf w}] =  
\sum_a  {1 \over {\rho k_a}} \,\,\oint_{\Gamma_a} d{\bf x}_a \cdot 
\left ( {{\delta F} \over {\delta {\bf x}_a}} \wedge 
{{\delta G} \over {\delta {\bf x}_a}} \right ) \,\,\, ,
\label{SLPB}
\ee
which can be used in a consistent way provided both $F$ and $G$
have a linear dependence on $k_a$'s as the currents of $\cal A$.
Another derivation of Eq. (\ref{SLPB}) is described
in Appendix A, where we reformulate the LP brackets within the
Clebsch picture of fluids in such a way that the dependence on
the diffeomorphism action is expressed explicitly.

A simple way to test the validity of the brackets just obtained
consists in checking whether they reproduce correctly the equation
of motion for the vortex filament by calculating explicitly the right
hand side of ${\partial_t {\bf x}} = \{ {\bf x} ,H \}$. The effect of
limit (\ref{LIM}) on $H$ and $\bf U$ is that of
exhibiting them into the form 
$$
H[{\bf w}_*] =
{\rho \over 2} \sum_a \sum_b \,\,k_a k_b
\oint_{\Gamma_a} \oint_{\Gamma_b} \,\, 
{ { d{\bf x}_a \, \cdot\, d{\bf x}_b} \over 
{4 \pi \vert {\bf x}_a - {\bf x}_b \vert }} \,\,\,,
$$
and
$$
{\bf U(x)} = \sum_a\,\,k_a\, \oint_{\Gamma_a}
\frac{d{\bf x}_a}{4 \pi \vert {\bf x} - {\bf x}_a \vert} \, \, ,
$$
respectively. Considering the single string case one finds
\be
\{ {\bf x} ,H \} \equiv {{\delta H} \over {\delta {\bf x}(s)}} 
\,\,\wedge \,\, {{{\bf {\dot x}}(s)} \over \rho} \, ,
\label{SEM}
\ee
where the functional derivative of $H$ is given by
$$
{{\delta H} \over {\delta {\bf x}(s)}} =
- {{ k \rho} \over {4 \pi}} 
\oint_{\Gamma} \, d{\bf y} \, \cdot  \! \int_{0}^L \! dr \, 
{{\delta } \over {\delta {\bf x}(s)}} 
{ {{\dot {\bf x}}(r)} \over {\vert {\bf x}(r) - {\bf y}\vert }}
$$
$$
=\rho \,\,{\dot {\bf x}}(s) \wedge  
k \oint_{\Gamma} {{({\bf x }(s) - {\bf y}) \wedge d{\bf y}} \over 
{4 \pi \vert {\bf x}(s) - {\bf y}\vert^3}} \,\,\, .
$$
Then, the expected equation of motion
$$
{\partial_t{\bf x}}
= \lambda(\Gamma) {\dot {\bf x}}+
k \oint_{\Gamma} { {({\bf x} -{\bf y}) \wedge d{\bf y}} \over 
{4 \pi \vert {\bf x} - {\bf y}\vert^3}}
= \lambda(\Gamma) {\dot {\bf x}}
- {\bf V({\bf x} ; \Gamma} )\, ,
$$
with $\lambda(\Gamma):= {\dot {\bf x}} \cdot{\bf V({\bf x};\Gamma})$, 
is achieved by explicitly calculating the wedge product in
Eq. (\ref{SEM}).
Notice that the component $\lambda(s,\Gamma)$ generates
displacements of $\Gamma$ that are parallel to $\Gamma$
itself due to its longitudinal character. The above result
easily extends to the many-component case.

\section{fine structure of 2d Current Algebra}
For planar vortices the notation of CA formalism can be
simplified in view of the fact that
${\bf w} = {\rm w} {\bf e_3}$ , ${\bf a} ={\rm curl} \,(A {\bf e_3}) =
\nabla A \wedge {\bf e_3}$ with $\nabla = {\bf e_1} \! \partial_{x_1}
+{\bf e_2} \! \partial_{x_2}$.
In particular, brackets (\ref{PBW}) reduce to
\be
\{ F ,G \}[\rm w] = {1 \over \rho} \int \! d^2x \,\,
{\bf w} \cdot \left ( \nabla \, {{\delta F} \over {\delta {\rm w} }}
\wedge \nabla \,{{\delta G} \over {\delta {\rm w} }} \right )\, ,
\label{DUE}
\ee
which, upon observing that a generic current is available in the two
forms  $J_{\bf a} [{\bf v}] = \rho \int d^2x \,\,{\bf a} \cdot {\bf v}
= \rho \int d^2x \,\,A\, {\rm w}\,\, = J_A [{\rm w}]$,
provides the current brackets
$$                                     
\{ J_A ,J_B \}[\rm w] = \rho \int d^2x \,\,
{\bf w} \cdot ( \nabla \,A  \wedge \nabla \,B ) =
\,J_{\{ A , B \}}[{\rm w}]
\, ,
$$
where
$\{ A , B \}_x
= {\bf e_3} \! \cdot \! (\nabla A \! \wedge \!\nabla \! B )$.

The two-dimensional LP structure just worked out can
be reformulated in such a way that the partition of
the ambient plane ${\bf R}^2$ in many sub-domains
is accounted for explicitly. This is realized through
the representation of the unit constant function
\be
1 = \sum_{a} \Theta_{a}({\bf x})
\label{UNO}
\ee
in terms of Heaviside functions
$\Theta_a({\bf x}) :=\Theta({\bf x}; S_a)$
nonvanishing inside the domain $S_a$.
The underlying idea is to show that implementing the fragmentation
process within the brackets formalism leads to recognize
the single components ${\rm w}_a ({\bf x})$ of $\bf w$ (associated to
plane domains $S_a$) as independent dynamical degrees of freedom.

The rule for selecting such domains is based on separating the
negative islands (where ${\rm w}< 0$) from the positive ones (where
${\rm w}> 0$). Such a situation indeed is usual since the condition
$\int d^2x\,{\rm w}({\bf x})=0$ 
is customarily assumed to exclude unphysical vortex configurations
whose energy cost is too high. On the other hand, the stable character
of such domains is ensured by the hydrodynamic laws of perfect fluids
which state the conservation of (the structure of) space patterns (the
partition in positive/negative vorticity domains, in the present
case) when the evolution is driven by area preserving diffeomorphisms.

Using Eq. (\ref{UNO}) any current $J_A [{\rm w}]$ can be
reexpressed
in terms of local currents as
\be
J_A [{\rm w}]
= \sum_{a} J_A^{(a)} [{\rm w}]
= \sum_{a} \rho \int_{S_{a}} d^2x \,\,A ({\bf x})\, {\rm w}({\bf x})
\, ,
\label{SCUR}
\ee
where $J_A^{(a)}[{\rm w}]
=\rho \int d^2x \,\,A ({\bf x})\, w({\bf x}) \Theta_{a}({\bf x})$.
In this way the additional information concerning the spatial
distribution of vorticity is explicitly taken into account
in the current description of the fluid. At the quantum level,
the quantity $J_A^{(a)}[{\rm w}]$ with $A \equiv 1$ are expected
to represent the quanta of vorticity
located in $S_{a}$. A simple calculation shows that the LP brackets
of local currents are given by
\be                   
\{ J_A^{(a)} ,J_B^{(b)} \} [{\rm w}] = \, \delta_{a b} 
J_{\{ A , B \}}^{a}[{\rm w}]
\label{BPB}
\ee
provided ${\rm w}_a({\bf x}) =0$ for $ {\bf x} \in \partial S_a$.
The vanishing of $\bf w$ on the boundary separating different
confining domains is crucial to eliminate the contributions coming
from the divergent character of
$$
\nabla \Theta_{a} ( {\bf x}) =
\oint_{\partial S_a } d {\bf y} \wedge {\bf e_3} \,
\delta^2 ({\bf x} - {\bf y} )
$$
on the boundary of $S_a$. This fact motivates as well the choice of
the set of plane domains $S_a$ based on distinguishing positive from
negative vorticity domains. Furthermore the LP brackets for any two
currents can be rewritten by means of the formula
\be
\{ J_A ,J_B \}[{\rm w}] =\! 
\Sigma_{a}\! \int_{S_{a}}
\!  \frac{d^2x}{\rho} \, {\bf w}_a \cdot \!
\left
( \nabla \, {{\delta J_A} \over {\delta {\rm w}_a}}
\wedge \nabla \,{{\delta J_B} \over {\delta {\rm w}_a}}
\right ) .
\label{NEW}
\ee
explicitly exhibiting the fine structure of the vorticity domains.
This also implies that $\{ {\rm w}_{a} := {\rm w}({\bf x}):\,
{\bf x}\in S_{a} \} $ can be considered as a set of fluid dynamical
variables.

It is worth noting how the fragmentation picture just introduced
allows one to enlarge the set of currents so as to include current
whose labels $A$ not necessarily vanish for $|{\bf x}| \to \infty$.
The case in which $A= x_1$, $B= x_2$ is illustrative of this.
From Eq. (\ref{BPB}) one readily obtains the canonical
coordinatelike brackets
\be
\{ J_{x_1}^{(a)} ,J_{x_2}^{(b)} \}[{\rm w}]
= \, \delta_{a b}  J_{1}^{(a)}[{\rm w}] 
= \delta_{a b} \,\rho K_{a} \, ,
\label{CANC}
\ee
where $K_a =\int_{S_a} \! d^2x \,{\rm w}({\bf x})$, that relate
the present field-theory description to the
pointlike vortex gas description of the Helmholtz standard model.
The quantum version of Eq. (\ref{CANC}) implies that the
information
related to the (average) position of the vortex domain $S_a$ on
the $x_1$-axis cannot be given together with that concerning the
position on the $x_2$-axis.
This clearly mimics the effects of the canonical quantization
rule standardly used for point vortices \cite{CHM} as well as the
uncertainty affecting the position of the string in the 3D case.
A nice magnetic-like interpretation of $J_{x_1}^{(a)}$,
$J_{x_2}^{(a)}$ is also available.
Rewriting first $x_1 {\rm w}({\bf x})$ ($x_2 {\rm w}({\bf x})$) in
$J_{x_1}^{(a)}$ ($J_{x_2}^{(a)}$) as
$$
x_r {\rm w}({\bf x})= {\bf v} \wedge {\rm curl}(x_r {\bf e_3})
-{\rm div} (x_r {\bf e_3} \wedge {\bf v} )\, , \, r= 1,2 \, ,
$$
and using then formula
${\bf e_3}\, {\rm div} ({\bf A} \wedge {\bf e_3}) \!
= \! {\rm curl}{\bf A}$
one finds $J_{x_2}^{(a)} \equiv {\bf e_1} \cdot {\bf P}_{(a)}$
and $-J_{x_1}^{(a)} \equiv {\bf e_2} \cdot {\bf P}_{(a)}$, where
$$
{\bf P}_{(a)} \equiv {\bf p}_{(a)} +
\oint_{\gamma_a} ( {\bf e}_3 \wedge
{\bf x})\, ({\bf v} \cdot d{\bf x}) \, ,   
$$
with ${\bf p}_{(a)} := \rho \int_{S_a} d^2x {\bf v} $ and
$\gamma_a = \partial S_a $. Such an expression makes visible
the structure of generalized magnetic moments characterizing
${\bf P}_{(a)}$ in which ${\bf p}_{(a)}$ represents the total
momentum pertaining to the domain $S_a$, while the circulation
term can be seen as an effective vector potential ${\bf A}
= B {\bf e_3} \wedge {\bf r}$, where
${\bf r} = \langle {\bf x} \rangle =
\oint \! (d{\bf x} \cdot {\bf v}) {\bf x} /K_a$ and $K_a$ is
the magnetic field $B$. Such a picture matches the magnetic
approach to the point vortex quantization presented in
Ref. \cite{PV}. 

Similarly to the 3D case, the Fourier mode algebra is obtained by
considering the Fourier decomposition
$A ({\bf x}) =\int d^2q \,\,A({\bf q}) e^{i{\bf q}{\bf x}}$
and defining the mode currents
\be
J_{\bf q}[{\rm w}] =
\int d^2 x \,\,{\rm w}({\bf x})\, e^{i{\bf q}{\bf x}} \, ,
\label{FMOD}
\ee
that represent the functionals whereby reconstructing any current as
illustrated by the formula
$J_A [{\rm w}] =  \rho \int d^2x \,\,A ({\bf x})\, {\rm w}({\bf x})
= \int d^2q \,\,A({\bf q}) J_{\bf q} [{\rm w}]$.
Moreover, the Poisson brackets of the $J_{\bf q}[{\rm w}]$'s are
readily derived from Eq. (\ref{BPB}) which provides the formula
$$
\{ J_{\bf q} ,J_{\bf p} \}[{\rm w}] = 
-{\bf e_3}\cdot ({\bf q} \wedge
{\bf p}) \, J_{\bf q+p} [{\rm w}] \; ,
$$
whereby its quantum mechanical counterpart
\be
[ J_{\bf q} ,J_{\bf p} ] [{\rm w}] = -i \hbar \,\,  
{\bf e_3} \cdot ({\bf q} \wedge {\bf p}) \,
J_{\bf q+p} ({\rm w}) \; ,
\label{BCO}
\ee
is derived. The resulting algebra coincides with the
well known algebra $W(\infty)$ (see, e.g., Ref. \cite{ZEIT}).

Now, going to the case of point vortices, it is interesting to
illustrate the diversity characterizing the scheme based on
the canonical variables and the procedure relying on the CA.
Quantizing a classical 2D vortex gas is usually
performed by replacing its classical Poisson brackets~\cite{SAF}
$\{x_a , y_b \} = \delta_{ab} / \rho k_a $ with the commutators
$[x_a , y_b ] =i \hbar \delta_{ab} / \rho k_a$ (see, e.g.,
Ref. \cite{CHM}). The definition of the ${\bf q}$ currents for
a pointlike vorticity distribution ensues directly from
Eq. (\ref{FMOD})
$$
J_{\bf q} (w) = \int d^2x \,\,w({\bf x}) e^{i{\bf q}{\bf x}} 
= \sum_a k_a \, {\rm e}^{i{\bf q}{\bf x}_a} \, ,
$$ 
where $ k_a = w({\bf x}_a)$, and local currents are recognized to
have the form $J_{\bf q}^a = e^{i{\bf q}{\bf x}_a}$. As a
consequence of the Baker-Campbell-Hausdorf formula
${\sl exp}(- {1 \over
2}[Q,P]) {\sl exp}Q {\sl exp}P = {\sl exp}(Q+P)$ it is found that  $$
e^{i{\bf q}{\bf x}_a}\,\,e^{i{\bf p}{\bf x}_a} 
e^{i \Phi_a(q,p)} =
e^{i({\bf p}{\bf x}_a + {\bf q}{\bf x}_a)} \, .
$$
\noindent
The phase $\Phi_a(q,p) =
({\hbar}/{2 \rho k_a}){\bf e_3}\cdot({\bf q}
\wedge {\bf p})$ is the nontrivial effect deriving from the
canonical quantization. In fact, while the commutator of two
any local currents still generates a current
(see Eq. (\ref{BPB})) since
\be
[ J_{\bf q}^a ,J_{\bf p}^b ] = \delta_{ab} \,\,2i\,k_a\,\, 
{\rm sin} [\Phi_a(q,p)] \,  
J_{\bf p + q }^a \quad,
\label{LCC}
\ee
the attempt to reconstruct the CA, namely the commutators
(\ref{BCO}), fails due to the nonlinearity of the sine factor
arising in Eq. (\ref{LCC}) that prevents the superposition of
local current $J_{\bf p + q }^a $ with different label $a$.
The usual result is recovered however either in the limit
$\hbar \rightarrow 0$ or when $k_a \rightarrow \infty$, both
entailing a semiclassical picture of vortices.

On the other hand, writing explicitly the current commutator
$$
[ J_{\bf q} , J_{\bf p} ] = 2i \sum_a \,k_a\,\, 
{\rm sin} [\Phi_a(q,p)] \, J_{\bf p + q }^a \, 
$$
shows the presence of an underlying magnetic-like structure
where two generators of planar displacements (magnetic translations)
commute provided the area element in the mode space
${\bf e_3}\cdot({\bf q} \wedge {\bf p})$ is equal (up to a factor
$\pi$) to the multiple fluxon $n\, ({2 \rho k_a}/{\hbar})$,
$n \in {\bf N}$. Also, it is worth noting that the structure
(\ref{BCO}) is partially recovered, namely
\be
[ J_{\bf q} ,J_{\bf p} ] = 2i\,k \, 
{\rm sin} [\Phi(q,p)] \sum_a  J_{\bf p + q }^a
\quad,
\label{REC}
\ee
when assuming the standard (low temperature) vorticity configuration
$|k_a|=k \equiv \hbar/m$ (due to the Feynman-Onsager condition, where
$m$ is the Helium atomic mass), for each
vortex. In this case local currents share the same sine factor which
keeps memory of the pointlike form of vortex cores.
A similar portrait has been depicted in Refs.~\cite{ARK}
and \cite{HOP} by considering certain realizations of the algebra
diff$(T^2)$ and su(N) and their link via contraction. More
precisely, su(N) has been shown to be equipped with commutators
whose structure constants have the form
$N {\rm sin} [{\bf e_3}\cdot({\bf q}\! \wedge\! {\bf p})
/N]$ which
reproduces that of diff$(T^2)$ for $N \to \infty$. 
%
\section{conclusions}
Based on a heuristic approach, we have shown in Sec. III that the
LP structure of string vortices can be evinced by combining the
effect of the string limit (\ref{LIM}) on currents and the request
that the algebraic structure of CA is preserved. Such a heuristic
way bypasses the complexity of Dirac's formalism.

A more detailed procedure has been supplied in Appendix A based
on the Clebsch potential picture of perfect fluids and the
explicit use of diffeomorphisms as the dynamical variables
in terms of which reformulating the LP brackets.

The applications of formula (\ref{SLPB}) are at least twofold.
First it is a crucial ingredient in constructing the functional
operator form of the currents of $\cal A$
in the implementation of the geometric quantization scheme
\cite{GMS2}, \cite{PRS} for string vortices.
Second, one can take advantage from formula (\ref{SLPB}) to study
the algebraic structure of string functionals such as Chen iterated
path integrals~\cite{CHEN} that represent the higher order topological
charges of the string~\cite{PRS1}.

The limiting process reveals the possible many-component
structure of the string. Such an aspect is fully accounted in the
planar vortex case discussed in Sec. IV and is used to make evident
the fine structure of the current algebra. Relying on such a
formulation of the dynamical algebra, we have reconstructed the CA
for (planar) point vortices showing how the canonical quantization
process yields a different algebraic structure for the local currents.
The difference disappears upon passage to an appropriate semiclassical
limit.

This effect is of course explainable as the manifestation of the
structural inequivalence between a model with a discrete distribution
of the vorticity and a (smooth) vorticity field theory. It might
be used in an explicit way to characterize the transition from
the (low temperature) rarefied gas of point vortex pairs to
a fluid with many interacting vortices, which takes place in
planar superfluids when temperature is raised. The many-vortex
fluid induces a more intense vortex interaction which possibly
requires a
fieldlike description capable of describing vortex
cores which are no longer reducible to pointlike objects.

\vskip 1. truecm
\centerline{\bf Acknowledgements}
\vskip 0.5 truecm
\noindent
The authors gratefully thank A. P. Balachandran, F. Lizzi, G. Marmo,
and P. W. Michor for stimulating discussions. The contribution
to this work of V. P. was mainly developed while he was visiting the
Erwin Schr\"odinger Insitute (E.S.I.) in Wien, Austria. He is grateful
to the E.S.I. for its hospitality and for supporting his visit.
The I.N.F.M. and the M.U.R.S.T. (within project Sintesi) are also
acknowledged for financial support.

\begin{appendix}
\section{}

The use of diffeomorphisms ${\bf x} \to {\bf y} =
{\bf \eta}({\bf x})$ as the dynamical variables of the system is
based on the Clebsch potentials (CP) picture of the vorticity
field ${\bf w}$. Within the CP picture assigning the field ${\bf v}$
(or ${\bf w}$, {\sl i.e.}, the state of the fluid) is equivalent
to defining the set of CP
\bq
\left \{ \left ( {\cal U}_j , \left ( \alpha_{j} , \beta_{j} ,
\varphi_{j} \right ) \right ) \, \bigl | \, {\bigcup}_j {\cal U}_j
= {\bf R}^3
\right \} \; , \nonumber
\eq
on a suitable covering of ${\bf R}^3$, such that
${\bf v}\equiv \!
k (\alpha_j \nabla \!\beta_j \! +\!\nabla\! \varphi_j )$
and $ {\bf w} \equiv k \nabla \alpha_j  \wedge \nabla \beta_j$
in the chart ${\cal U}_j$ (index $j$ referred to local charts
is dropped in the sequel to simplify formulas).
Triads of CP provide an alternative system of coordinates
represented by the map (with its inverse)
\bq
\left ( \alpha ({{\bf x}}) , \beta ({{\bf x}}) , \varphi ({{\bf x}})
\right ) \,
{{{~~~}\atop{\longrightarrow}}\atop{{\longleftarrow}\atop{~~~}}}
\, {{\bf x}} = {{\bf x}} \left ( \alpha , \beta , \varphi \right ) 
\nonumber
\eq 
whose definiteness is ensured by the fact that its Jacobian
\bq
I \left ({\bf x}\right )   =
\nabla \varphi \cdot \left (\nabla \alpha \wedge \nabla \beta \right )
= \frac{{\bf v} \cdot {\bf w}}{k^2} 
\label{JAC}
\eq
is nonvanishing
(namely the topological charge is nonzero \cite{KUMI}).
The Jacobian furnishes further geometric information.
In particular, a set of six equations can be easily worked out
from Eq. (\ref{JAC}) two of which read
\bq
I\, \frac{\partial {\bf x}}{\partial \varphi} =
\nabla \alpha \wedge \nabla \beta \; , \;
\nabla \varphi  = I \, \frac{\partial {\bf x}}{\partial \alpha}
\wedge \frac{\partial {\bf x}}{\partial \beta}
\; ,
\label{ABC}
\eq
while the others are obtained by cyclic permutations of $\alpha$,
$\beta$, $\varphi$.
The ${{\bf w}}$ can be thought of as the set of its {\it fibers}
(that is its integral curves) filling the whole ambient space
${\bf R}^3$. Fibers ${{\bf x}}\left (\alpha ,\beta ,\varphi \right)$,
in turn, embody the topological structure of ${\bf w}$ and are
homotopic to each other (the extended version of such a review
can be found in Ref. \cite{PRS}). 

The time evolution involves the change driven by the
time-dependent diffeomorphisms $\eta$
$$
({\it a}({{\bf x}}) , {\it b}({{\bf x}}) , {\it f}({{\bf x}}))
\to
(\alpha ({{\bf x}}) , \beta ({{\bf x}}) , \varphi ({{\bf x}})) \, ,
$$
where
$\alpha \! ({\bf x}) := {\it a} [{\bf \eta} ({\bf x})]$,
$\beta \! ({\bf x}) :=  {\it b} [{\bf \eta} ({\bf x})]$,
$\varphi \! ({\bf x}) := {\it f} [{\bf \eta} ({\bf x})]$,
and ${\bf \eta}({\bf x}) \equiv {\bf x}$ at the initial time.
This allows one to regard ${\bf y} ={\bf \eta}({\bf x})$ as
a dynamical variable.
The kernel of formula (\ref{PBW}) for $F= J_A$, $G =J_B$
reduces to ${\bf w} \cdot ({\bf a} \wedge {\bf b})$
which represents the result we must reproduce by introducing
${\bf y}$-dependent functional derivatives.
Upon expressing a current as
\be
J_{\bf A} [{\bf w}] =
k \rho \int \! \! d^3x \,\,{\bf A}({\bf x}) \cdot
(\nabla \alpha\! ({\bf x}) \wedge \nabla \beta\! ({\bf x})) \, ,
\label{NPB}
\ee
where $\alpha\!$, $\beta$ contain ${\bf \eta}$, the
functional derivative of $J_A$ can be written through the formula 
\be
{{\delta J_A} \over {\delta y^k }}\, \nabla y^k 
=\rho {\bf w}({\bf x}) \wedge {\bf a} ({\bf x}) := D_y J_A \, .
\label{PRO}
\ee
The identity ${\bf w} \cdot ({\bf a} \wedge {\bf b})
=({\bf w}/{\bf w}^2 ) \! \cdot\!
\left [ D_y J_A \,  \wedge \, D_y J_B \right ]$,
where ${\bf w}=
\nabla \alpha\! ({\bf x}) \wedge \nabla \beta\! ({\bf x})$
must be considered as dependent on $y^k= \eta^k ({\bf x})$,
is derived from Eq. (\ref{PRO}). Then the brackets
\be
\{ J_A , J_B \}({\bf w}) = {1 \over \rho} \int d^3x \,\,
\frac{ {\bf w} }{ {\bf w}^2 } \! \cdot\!
\left [ D_y J_A \,  \wedge \, D_y J_B \right ] \, ,
\label{CROS}
\ee
can be defined. Since $D_y J_A$ is also available in the form
\be
D_y J_A \equiv \rho\, I({\bf x}) \, \frac{d {\bf x}}{d \phi} \wedge
{\bf a} ({\bf x}) \, ,
\ee
one finds that the string limit of Eq. (\ref{CROS}) is well
defined in that the factor $1/{\bf w}^2$ generating a divergent
contribution is compensated by two factors $I({\bf x})$ coming
from $D_y J_A$ and $D_y J_B$. In the string limit, in fact, one
can easily
show that the approximation $x_3 \simeq \phi$ can be locally
implemented around the vortex core as a consequence of the
confinement of vorticity inside a thin cylinder.
In view of Eq. (\ref{ABC})
this implies that $I({\bf x}) \simeq |{\bf w}({\bf x})|$.
Consequently, one finds
$$
\frac{D_y J_A}{\rho |{\bf w}|} = \frac{d {\bf x}}{d \phi} \wedge
{\bf a} ({\bf x}) \to
\left [{{\delta J_A} \over {\delta {\bf x}}} \right ]_{\Gamma}
$$
where the subscript $\Gamma$ recalls that ${\bf x} \in \Gamma$
and $\phi$ identifies with the
arclength in the above expression.
\end{appendix}
%


\end{multicols}


\begin{references}
%
%
\bibitem{DON}
R. J. Donnelly, {\it Quantized Vortices in He II},
(Cambridge University Press, Cambridge, England, 1991). 
%
\bibitem{BG2}
{\it Superfluidity and Superconductivity}, 2nd ed., edited 
by D. R. Tilley and J. Tilley (Hilger, London, 1986). 
%
\bibitem{BEC} F. Dalfovo,
S. Giorgini, Lev P. Pitaevskii, and S. Stringari,
Rev. Mod. Phys. {\bf 71}, 463 (1999).
%
\bibitem{NEW} Beyond various references listed below
a pair of recent papers on this topic are
U. R. Fischer, Ann. Phys. (NY) {\bf 278}, 62 (1999);
J. P. Kottmann and A. M. Schakel,
Phys. Lett A, {\bf 242}, 99 (1999).
%
\bibitem{RR} M. Rasetti and T. Regge, Physica A {\bf 80}, 217 (1975).
%
\bibitem{GMS1} G. A. Goldin, R. Menikoff, and D. H. Sharp,
Phys. Rev. Lett. {\bf 58}, 174 (1987). 
%
\bibitem{GMS2}
G. A. Goldin, R. Menikoff, and D. H. Sharp, Phys. Rev. Lett.
{\bf 67}, 3499 (1991).
%
\bibitem{PS1}
V. Penna and M. Spera, J. Math. Phys. {\bf 30}, 2778 (1989).
%
\bibitem{PRS}
V. Penna and M. Spera, J. Math. Phys. {\bf 33}, 901 (1992);
V. Penna, M. Rasetti, and M. Spera, Contemp. Math.
{\bf 219}, 173 (1998).
%
\bibitem{WU}
Y. Wu, J. Math. Phys. {\bf 34}, 2342 (1993).
%
\bibitem{PRS1}  V. Penna, M. Rasetti, and M. Spera,
Int. J. Mod. Phys. B {\bf 4}, 1289 (1990);
V. Penna and M. Spera,
{\it ibid.} B {\bf 6}, 2209 (1992).
%
\bibitem{FET}
A. L. Fetter,  Phys. Rev. {\bf 162}, 143 (1967).
%
\bibitem{BIGAS}
This becomes evident by observing that one of the three
coordinates required to describe a 1D curve can be always used
to parametrize locally the curve itself.
Setting $s \equiv x_k$ enables the reduction of
Eq. (\ref{SBR}) to the
canonical form $\{ x_i(s) , x_j(s') \}_* \, = \, \epsilon_{ijk} 
\delta(s-s') /{k \rho}$. 
It is worth noting that such a result makes the set of points
representing the intersections of $\Gamma$ with a given plane 
$x_k =$ const equivalent to a planar model of point vortices
with off-plane couplings.
%
\bibitem{KUMI} E. A. Kuznetsov and A. V. Mikhailov,
Phys. Lett. {\bf 77 A}, 37 (1970).
%
\bibitem{BG1}
{\it Proceedings of the 20th International Conference on
Low Temperature Physics}, edited by R. J. Donnelly,
[Physica B {\bf 194-196}, (1994)].
%
\bibitem{BG3} Vortex dynamics and Quantum effects
in Josephson junction arrays are reviewed in the Proceedings of the 
ICTP Workshop on {\it Josephson Junction Arrays},
edited by H. A. Cerdeira 
and S. R. Shenoy [Physica B {\bf 222}, 336 (1996)]. 
%
\bibitem{PV} V. Penna, Phys. Rev. B {\bf 59}, 7127 (1999).
%
\bibitem{ARN} V. I. Arnold,
{\it Mathematical Methods of Classical Mechanics}
(Springer-Verlag, Berlin, 1978).
%
\bibitem{ARK} V. I. Arnold and B. A. Khesin,
{\it Topological Methods in Hydrodynamics}
(Springer-Verlag, New York, 1998).
%
\bibitem{MWA} J. Marsden and A. Weinstein,
Physica {\bf D \, 7}, 305 (1983). 
%
\bibitem{RR2} M. Rasetti and T. Regge, in
{\it Highlights of Condensed Matter Theory},
edited by F. Bassani (Compositori, Bologna, 1985).
%
\bibitem{ZEIT} V. Zeitlin, Physica D {\bf 49}, 353 (1991).
%
\bibitem{SAF} P. G. Saffman, {\it Vortex Dynamics}
(Cambridge University Press,  Cambridge, England, 1992);
H. Aref, Annu. Rev. Fluid Mech. {\bf 15}, 345 (1983).
%
\bibitem{CHM} R. Y. Chiao, A. Hansen, and A. Moulthrop,
Phys. Rev. Lett. {\bf 55}, 2887 (1985).
%
\bibitem{HOP} J. Hoppe,
Int. J. Mod. Phys. A {\bf 4}, 5235 (1989).
%
\bibitem{CHEN} K.-T. Chen, Adv. Math. {\bf 23}, 181 (1977);
Bull. Am. Math. Soc. {\bf 83}, 831 (1977).
%
\end{references}
\end{document}